# AN INNOVATIVE APPROACH FOR E-GOVERNMENT TRANSFORMATION

Ali M. Al-Khouri

Emirates Identity Authority, Abu Dhabi, UAE.
ali.alkhouri@emiratesid.ae

## ABSTRACT

*Despite the immeasurable investment in e-government initiatives throughout the world, such initiatives have yet to succeed in fully meeting expectations and desired outcomes. A key objective of this research article is to support the government of the UAE in realizing its vision of e-government transformation. It presents an innovative framework to support e-government implementation, which was developed from a practitioner's perspective and based on learnings from numerous e-government practices around the globe. The framework presents an approach to guide governments worldwide, and UAE in particular, to develop a top down strategy and leverage technology in order realize its long term goal of e-government transformation. The study also outlines the potential role of modern national identity schemes in enabling the transformation of traditional identities into digital identities. The work presented in this study is envisaged to help bridge the gap between policy makers and implementers, by providing greater clarity and reducing misalignment on key elements of e-government transformation. In the hands of leaders that have a strong will to invest in e-government transformation, the work presented in this study is envisaged to become a powerful tool to communicate and coordinate initiatives, and provide a clear visualization of an integrated approach to e-government transformation.*

## 1. INTRODUCTION

Among the many promises of the Information Communication Technologies (ICT) revolution is its potential to modernise government organisations, strengthen their operations and make them more responsive to the needs of their citizens. Many countries have introduced so-called e-government programmes that incorporate ICT and use it to transform several dimensions of their operations, to create more accessible, transparent, effective, and accountable government [1-3]. In recent years, e-government development has gained significant momentum despite the financial crisis that crippled the world economy [5-6]. For most governments, the recent financial crisis was a wakeup call to become more transparent and efficient [7]. In addition, there is also growing demand for governments to transform from a traditional agency and department centric model to a "Citizen-Centric" model [8-10]. Such a transformation is expected to enhance the quality of life of citizens in terms of greater convenience in availing government services [1] and thereby result in increased customer satisfaction levels and trust in government [11-13]. Government agencies are increasingly embracing Information and Communications Technology (ICT) to boost efficiency and integrate employees, partners and citizens in a seamless manner [14,15]. On the other hand, it is becoming increasingly difficult to achieve these outcomes and meet the needs of the citizens with fragmented e-government initiatives (ibid). Such a situation is forcing many governments to take an integrated approach to improve the effectiveness of delivering services to their citizens [16]. Having closely studied many of the leading e-government programs around the world, some of which have formed dedicated e-government institutions to deliver the desired transformation, we see that very few have succeeded in achieving the outcomes they initially hoped to deliver. This does not mean that the world has not witnessed any e-government initiatives that have succeeded in delivering effective e-services to citizens. Rather we see that most e-government programs have automated and digitized some existing processes rather than transformed government services. E-government is not just about enabling existing government services on the Internet, but rather is about a re-conceptualization of the services offered by governments, with citizens' expectations at the core of the re-conceptualization. As such, this can only be achieved through vertical and horizontal integration of government systems to enable communications crossing the boundaries of the different government agencies and departments, which should result in a "one stop service centre" concept. The existing body of knowledge is full of strategies, frameworks, and approaches, developed by consulting companies or by academic researchers, however, practitioners in the field of government have been hesitant to accept or fully believe in the practicality of these frameworks.

This research study was particularly developed to provide an analysis of the current e-government status-qua in the United Arab Emirates and to support the government in pursuing its objective towards e-government transformation. Thus, it offers an innovative framework from a government practitioner's viewpoint and in light of the existing literature in the field. The recommended framework is an amalgamation of learnings from various e-governments initiatives across the globe. It defines a comprehensive approach addressing technology, strategy and the broader approach to realizing e-government transformation. It proposes many innovative models to support the visualization of numerous dimensions of transformed e-government. This research article is structured as follows. First, a short literature review on the concept of citizen centricity in e-government applications is provided. Next, some recent statistics on the progress of e-government with focus on the UAE is presented, covering some of the recent efforts of the government of the UAE in terms of its strategy, e-services and distribution, and recent developments. The research and development methodology is outlined thereafter, and subsequently the proposed framework is presented and discussed. The paper concludes with the presentation of some key thoughts and considerations around success factors and improvement opportunities

## 2. DEPARTMENT CENTRIC TO CITIZEN CENTRIC

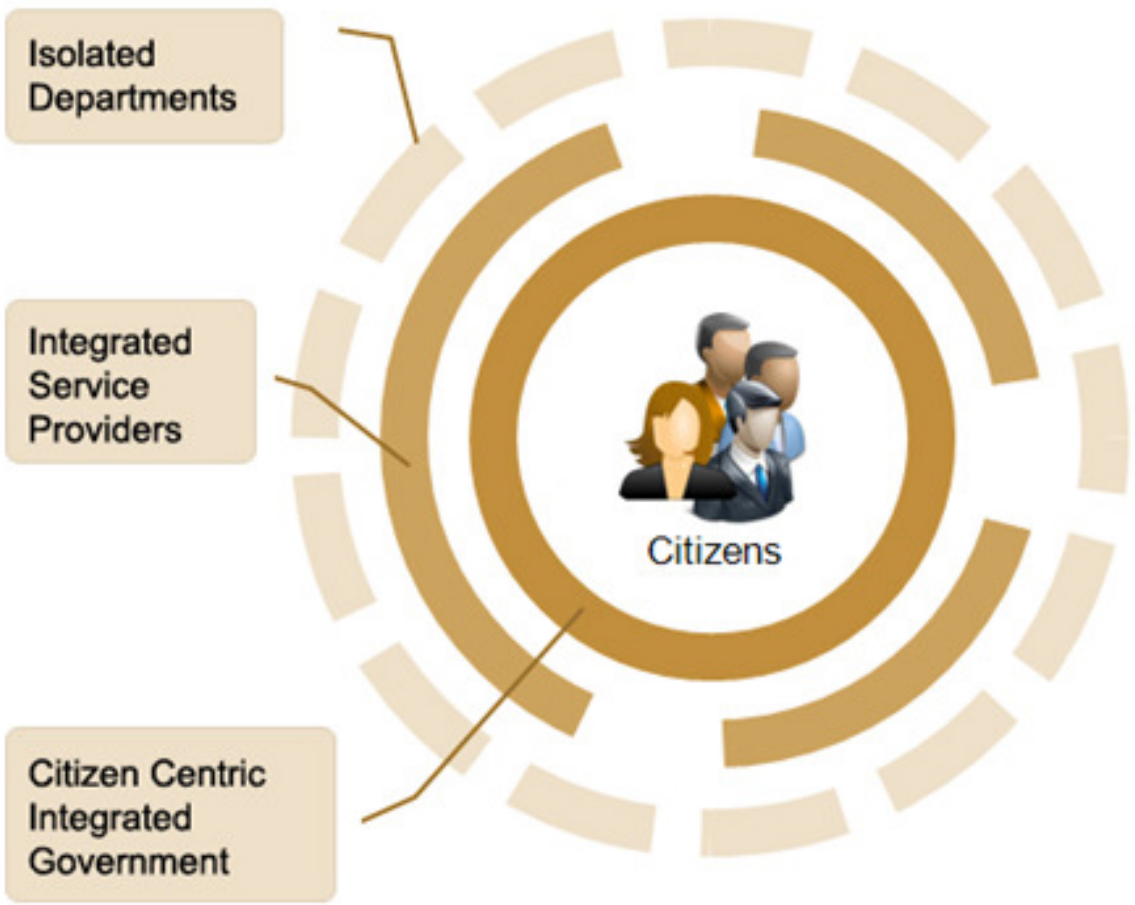

Figure-1: Stages of e-government transformation

Modern governments are steadily transforming from the traditional department centric model to a citizen centric model for delivering services [8-10,17]. Such a model aims to change the perspective of government constituents, so that they view their government as an integrated entity rather than discrete agencies and departments. However such a transformation has multiple stages. Figure-1 attempts to depict that a Government consists of various agencies and service providers each of which has many departments offering services to citizens. In a department centric approach, citizen needs to interact with each department separately causing inconvenience and inefficiency. Moreover, any services that requires approvals or intervention of more than one department, would take a long time to deliver. The next stage in the transformation process is the integration at the service provider level where multiple services and departments under a single agency or service provider are integrated to give a single agency feel to the citizens. However, citizens still need to interact with different agencies for different purposes, leading to less transparency and convenience for the citizen. Fully integrated government provides vertical[1] and horizontal[2] cross service providers and cut through various layers of delivery [12]. Government integration results in projecting a single government view to the citizen and allows them to avail services from One-Stop-Shop portals and Service Access Points.

## 2.1. Characteristics of Citizen Centric E-government

Citizen centric e-government should (or would) enjoy increased trust of citizens and should ensure accountability of government transactions [18]. It should also provide enhanced collaboration among departments and stakeholders, thereby enabling fast decision making and consensus [19]. Citizen centric e-government could also help avoid duplication and overhead through shared services and infrastructure, thereby helping achieve reduced service delivery cost while enhancing customer satisfaction. Business intelligence gathered via integrated service provision would also enable the government to track the effectiveness of initiatives and schemes and enhance decision making. Citizen centric e-government in its final form would provide improved transparency and consistent user interfaces and convenient channels for citizens to access e-government services [20]. Via the enforcement of strict Service Level Agreements (SLA) with all government entities, government can ensure that citizens get improved responsiveness for their service requests and increased security and privacy, thereby earning their trust when they avail e-services [21]. In addition, these services are also available anywhere, and anytime, breaking the traditional limitations of public sector working hours. Effective e-government integration would provide opportunities for businesses to provide inputs and to air concerns, increase transparency and a serve to level playing field for service offerings [22]. Businesses should stand to gain from faster clearances of permits and licenses, reduced overhead, improved customer service and verification of customer identities in a fast and reliable manner. The next section will shed light on some recent statistics on e-government progress worldwide, with a specific focus on the progress made by the UAE.

# 3. E-GOVERNMENT WORLDWIDE AND IN THE REGION

UN agency known as UN Public Administration Network (UNPAN) benchmarks global governments against four key metrics – 'Online Service Index', 'Telecommunication Infrastructure Index', 'E-Participation Index' and 'Human Capital Index'. These indices collectively represent measurements of a nation’s readiness in terms of (1) telecommunication infrastructure, (2) maturity of e-services, (3) participation of citizens in decision making and (4) human resource availability to meet the requirements of offering e-government services. Below are highlights of the most recent UN survey conducted in 2010 [23].

[1] *Vertical Integration:* This stage initiates the transformation of government services rather than automating its existing processes. It focuses on integrating government functions at different levels, such as those of local governments and state governments.
[2] *Horizontal Integration:* This stage focuses on integrating different functions from separate systems so as to provide users a unified and seamless service.

## 3.1. UN E-government Survey 2010 Findings

Table-1 above lists the top 10 countries in the UN survey 2010 [*see also* 24]. As per the UN report these countries have achieved maturity in the transactional stage of e-government. For example The Republic of Korea has consolidated its position in offering transactional e-services and is planning to achieve transformation towards citizen-centric e-government by the year 2012. Table-2 provides the UN rankings of the top six Middle East countries in terms of their e-government readiness over a 5 year period. Of the six countries listed above, five are from the GCC, with only Jordan being a non GCC Arab country included. Bahrain and the UAE were ranked top two respectively, followed by Kuwait at 3rd, Saudi Arabia at 5th, Qatar at 6th and Oman at 8th. It is also observed that there has been steady progress by Bahrain in the field of e-government. Between 2008 and 2010, Bahrain has made remarkable progress in terms of improving its UN e-government ranking, jumping 29 points up to rank 13 worldwide after being ranked at 42 in the UN eGovernment Readiness Survey in 2008. Saudi Arabia has also advanced from (70) to (58), and Kuwait from (57) to (50), and Oman from (84) to (82), which is attributed, in general, to these countries further investment in IT infrastructure. On the other hand, the UAE fell 17 points, slipping from 32nd to 49th while Qatar fell by nine ranks, moving from 53rd to 62nd.

Table 1: Top 10 Countries in the UN Survey 2010

| **Rank** | **Country** | **Index** |
|---|---|---|
| 1 | Republic of Korea | 0.8785 |
| 2 | United States | 0.8510 |
| 3 | Canada | 0.8448 |
| 4 | United Kingdom | 0.8147 |
| 5 | Netherlands | 0.8097 |
| 6 | Norway | 0.8020 |
| 7 | Denmark | 0.7872 |
| 8 | Australia | 0.7864 |
| 9 | Spain | 0.7516 |
| 10 | France | 0.7510 |

Table 2: UN Ranking of Middle East Countries e-Gov Readiness

| **Country** | **2010 Ranking** | **2008 Ranking** | **2005 Ranking** |
|---|---|---|---|
| Bahrain | 13 | 42 | 53 |
| UAE | 49 | 32 | 42 |
| Kuwait | 50 | 57 | 75 |
| Jordan | 51 | 50 | 68 |
| Saudi Arabia | 58 | 70 | 80 |
| Qatar | 62 | 53 | 62 |

Though one may view these findings as being a numbers game, the e-government index provides governments the opportunity to look little deeply into their long-term strategy and the short-term policy for quick performance. Overall, the survey pointed out that e-government initiatives in GCC countries have helped underpin regulatory reform, while promoting greater transparency in government. The survey results are hoped to play a key role in enhancing the delivery of public services, enabling governments to respond to a wider range of challenges despite the difficulties in the global economy.

## 3.2. E-government Progress in UAE

The UAE has been at the forefront of adopting advanced technologies to improve the efficiency of governance. The visionary leadership of UAE has initiated numerous e-government programs aimed at enabling the government in effective policy making, governance and service delivery. A key focus of the 2011-2013 UAE Government strategy is to improve government services and bring them in line with the international standards, with special emphasis on education, healthcare, judicial and government services. The principles of the UAE e-government strategy are summarized as follows:

- Maintain continuous cooperation between federal and local authorities;
- Revitalize the regulatory and policy making role of the ministries, and improve decision making mechanisms;
- Increase the efficiency of governmental bodies, and upgrade the level of services by focusing on customer needs;
- Develop civil service regulations and human resources, by focusing on competence, effective Emiratization and leadership training;
- Empower Ministries to manage their activities in line with public and joint policies;
- Review and upgrade legislations and regulations

The UAE has been going through various stages of e-government developments. In order to provide a clearer perspective of where the UAE stands against international benchmarks, collectively as a nation and as individual emirates, the below data has been used to provide a common understanding of the status quo. The data samples considered for this exercise are indicative but not exhaustive, and have been compiled based on publicly available information.

### 3.2.1. E-Services Profile of UAE

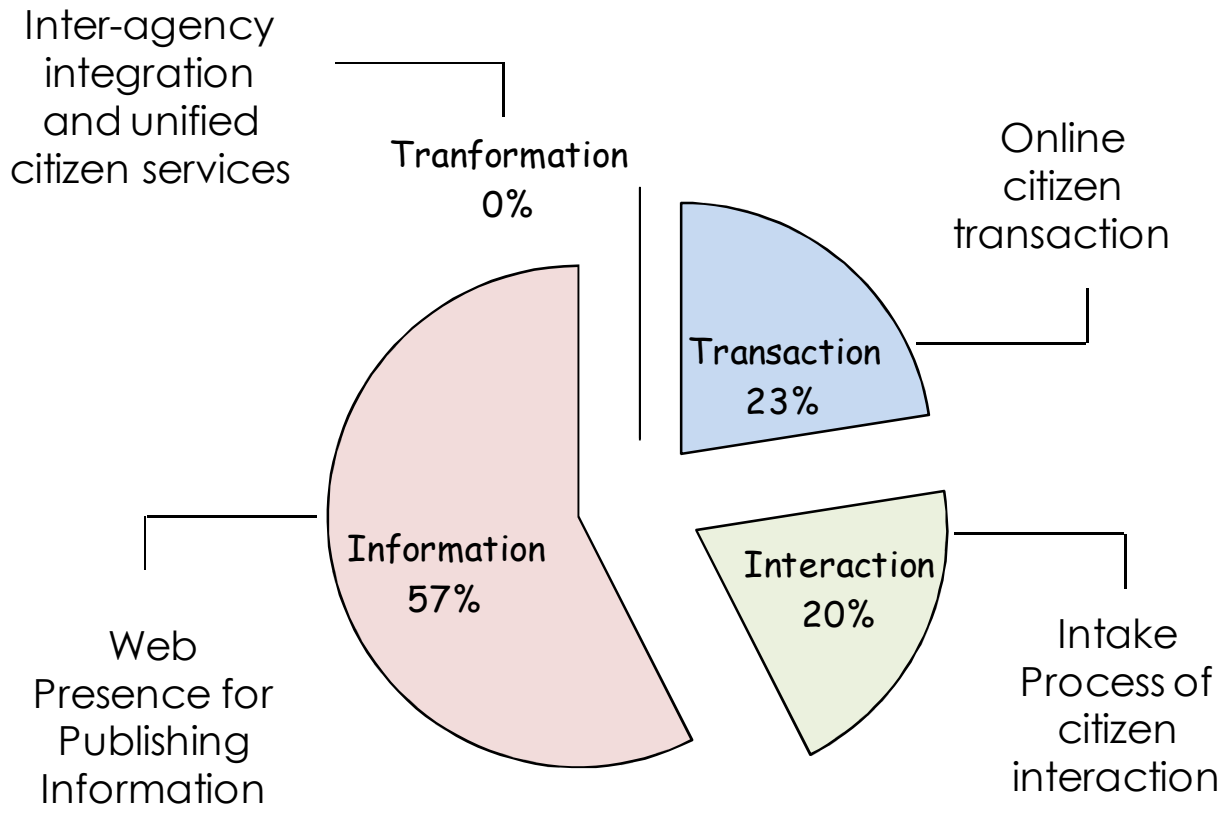

**Note:** Figures are based on limited sample of publicly available information and is only indicative

Figure-2: UAE e-government profile summary

Layne and Lee [25] developed a four-stage process to depict the e-government applications evolvement. These are Information, Interaction, Transaction and Transformation. The first stage embraces the publication of information on websites for citizens seeking knowledge about procedures governing the delivery of different services. The second stage involves interactivity where citizens can download applications for receiving services. The third stage involves electronic delivery of documents. The fourth stage involves electronic delivery of services where more than one department may be involved in processing a service request or service. The following chart (Figure-2) depicts a summary of the status of various services in the UAE as benchmarked against the commonly used stages of e-government i.e., Information, Interaction, Transaction and Transformation. From the above data, it is evident that collectively as a nation, the UAE government e-services are at the 'Information' stage and there is an equal distribution of e-Services between 'Interaction' and 'Transaction' stages. The important observation to be noted is that there is bigger challenge of inter-agency integration (Ready, 2004), which is the key to achieving 'Citizen-Centric' e-government.

### 3.2.2. E-Services Distribution Across Emirates

Having seen the overall e-government status across UAE, the following Figure-3 shows the distribution of government e-services across each emirate. From the graph it is evident that the e-government initiatives in Abu Dhabi and Dubai are more advanced than the other emirates and have the foundation for providing citizen-centric services. Based on this foundation there is growing momentum at the federal level to move towards shared services and increased integration.

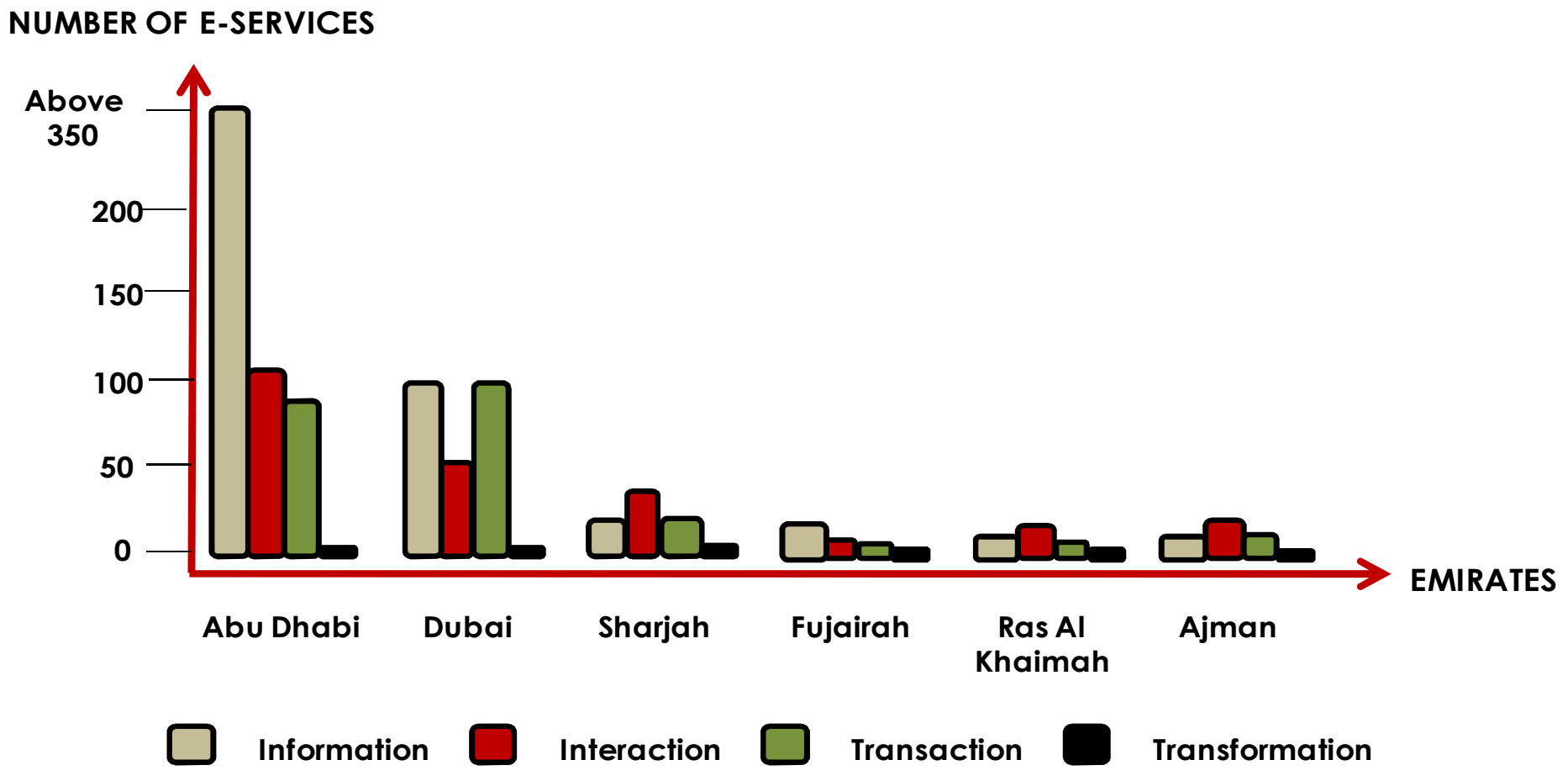

**Note:** Figures are based on limited sample of publically available information and is only indicative. The figure are high in Abu Dhabi is because most of the federal ministries are based in Abu Dhabi.

Figure-3: E-services distribution across Emirates

### 3.2.2. Current Stage of UAE in E-government Evolution

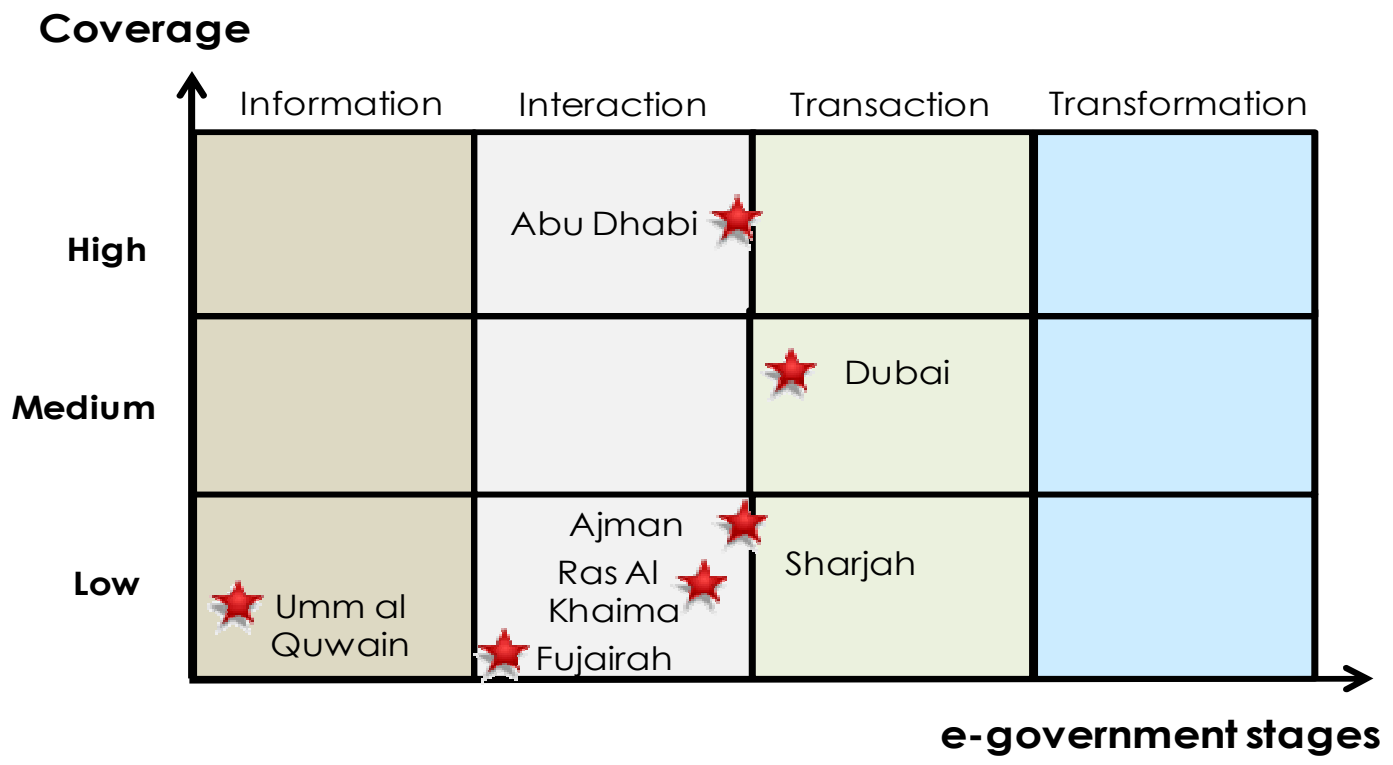

**Note:** e-government stage across the UAE is observed to be transitioning from interaction to transaction.

Figure-4: e-Government stages across UAE

Figure-4 illustrates the stage of each emirate in the UAE and the evolution of e-government in each. Combining the observations made so far, we can infer that the UAE as a nation is in the transition stage from Interaction to Transaction. While service coverage (i.e., number of services) is higher in Abu Dhabi, Dubai has made more progress towards implementation of transactional services. The UAE government has been making resolute and strong progress

towards laying the fundamental infrastructure needed to enable the e-government environment. The UAE has one of the highest quality broadband connections in the world, according to findings by the University of Oxford [26]. According to a recent research published by the Economist Intelligence Unit, the UAE was found to be leading the Middle East region in terms of its continued and steady improvement in broadband, mobile and Internet connectivity levels [27]. See also Figure-5. On the other hand, the UAE is now considered to have the highest rates of fibre optic[3] penetration in the world, according to research carried out by IDATE on behalf of the FTTH Council Europe Middle East Working Group [28]. The UAE is ranked fourth in the world, with 30.8 per cent of the country's households and businesses connected to fibre optic systems, behind Japan, South Korea and Hong Kong. The UAE is described as representing 96% of the Middle East region's FTTH/H subscribers and 76% of all homes passed by fibre.

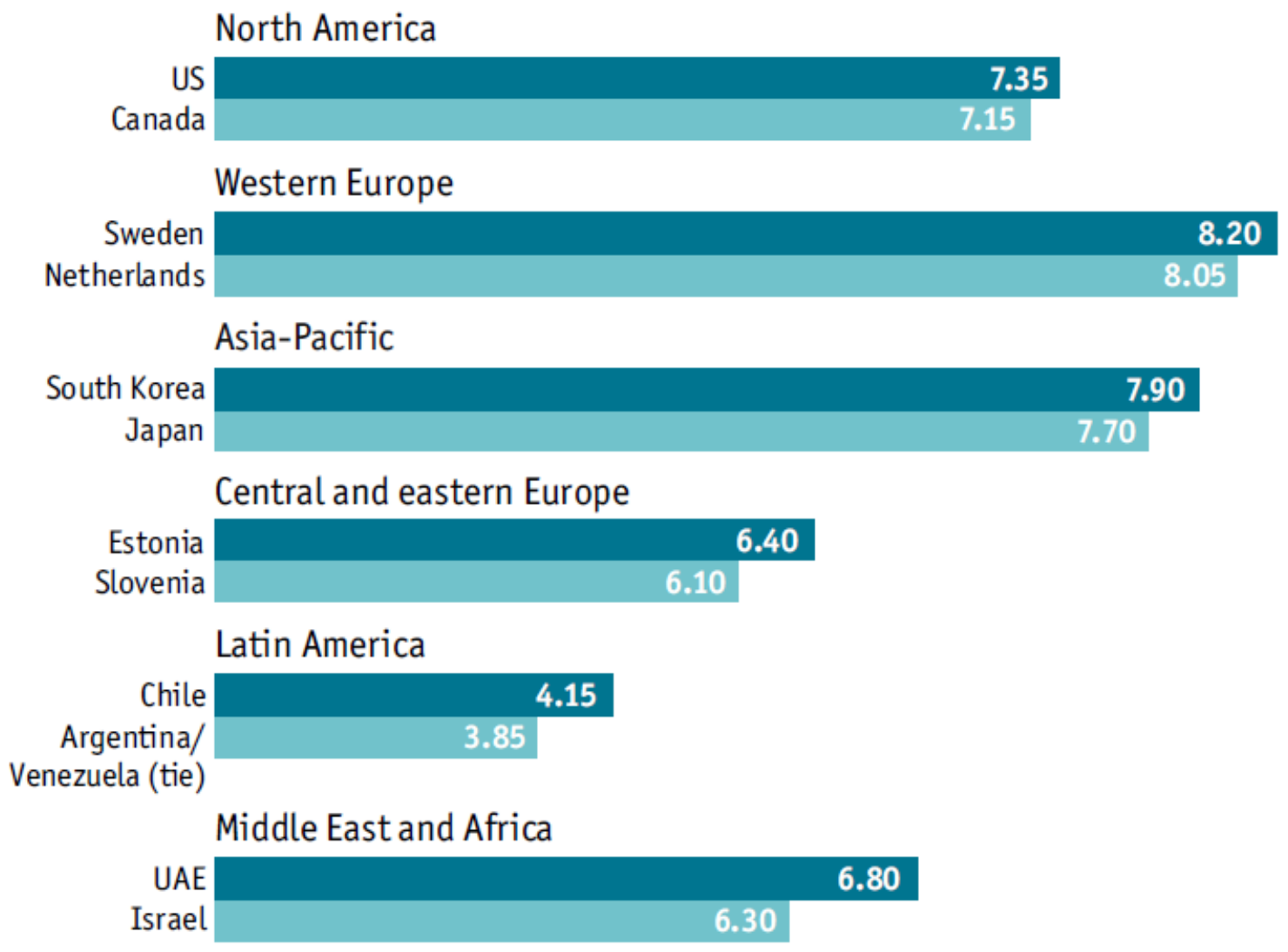

Figure -5: Regional digital economy rankings leaders: connectivity and technology infrastructure [27]

### 3.2.4. E-Government and the National Identity Management Infrastructure

To meet the growing need to integrate citizens into e-government initiatives, the smart citizen ID card initiatives adopted by many countries are meant to provide reliable methods for identifying and authenticating citizens availing e-services. An earlier research study in which the author participated and published in 2007 indicated that if essential components are integrated with such systems, such programs have the potential to address key challenges facing e-government initiatives, specifically those related to G2C [29-30]. The new National Identity Card Scheme rolled out in the UAE in 2005 is one of the largest federal government programs in the country to provide a cost effective, multi-functional, robust and secure national identity management infrastructure. The program is considered to mark a major milestone in the development of e-government; allowing citizens to authenticate themselves in an easy and comprehensively secure and electronic way whenever they access e-government applications.

The government announced recently the kickoff of a Public Key Infrastructure (PKI) and a Federated Identity Management (FIM) project to complement the existing identity management infrastructure and provide extended services to federal and local e-government authorities in the UAE [31]. The project aims to develop a comprehensive and integrated security infrastructure to enable a primary service of confirmed digital identities of UAE ID card holders on digital networks; primarily on the internet. The project has two strategic objectives: (1) to enable verification of the cardholder's digital identity; (authentication services) by verifying PIN, biometric, and signature and (2) provide credibility (validation services) through the development of a Central Certification Authority. PKI is regarded as a crucial component to provide higher security levels in digital forms, and may have a multiplier effect if integrated with the existing government trusted identity management systems. To support and enhance this capability many folds, this research study puts forward an innovative framework, referred to here as CIVIC IDEA, an abbreviation for *"**C**itizen **I**nclusive **V**ision realized through **ID** **C**ard **I**ntegrated **D**elivery of **E**-government **A**pplications."* The approach is envisaged to support the government of the UAE in achieving its vision of e-government transformation, while leveraging the strengths of the UAE national ID card initiative i.e., building upon the capabilities provided by the new smart ID card relating to the authentication capabilities of individuals over digital networks. The following section will shed light on the research and development methodology, and the proposed framework is discussed afterwards.

[3] Fibre optic connections enable almost unlimited volumes of digital data to be transmitted using pulses of light. The technology is replacing traditional copper wiring for broadband internet networks.

## 4. RESEARCH AND DEVELOPMENT METHODOLOGY

This research is more qualitative than quantitative in nature, although it relies on extensive analysis of case studies related to federal e-Government strategies through literature review. The analysis involved mapping of the federal e-Government strategies and the countries ranking in the overall e-Government index of UN survey with focus on the United Arab Emirates. This provided some thoughts related to what strategies could yield more successful results to enhance the UAE's position in the UN rankings. The study tried to balance the intensity of data collection of the case studies. Too many constructs could have led to a complex framework. Inadequate volume of data or sparse variation on the other hand might have failed to capture the whole picture in its entirety. The researchers were aware of these potential risks and worked to avoid them. Components and layout of the framework have converged from accumulated evidence (qualitative data). Gradually, a generic framework began to emerge. The researchers also compared the emergent framework with evidence collected from the multiple cases one at a time. We continued this iterative process until the data corroborated well the evolving framework. Finally, we consulted literature for contradiction or agreement. In many cases this helped form more perspectives. The following formatting rules must be followed strictly. This (.doc) document may be used as a template for papers prepared using Microsoft Word. Papers not conforming to these requirements may not be published in the conference proceedings.

## 5. PROPOSED FRAMEWORK

From the extensive literature review conducted and the analysis of the UN e-government survey reports, it was amply clear that the political leadership and e-government leaders need simpler and effective tools for visualizing and conveying the strategies. Based on this need, this study was focused on developing simplified models and tools for understanding and managing e-government initiatives. These models design containing key information resembles the issues and challenges faced by e-government initiatives that can then become the focal point around which decisions for business change and/or improvement of operations are made. Curtis [32] identified five different components that need to be considered in a modelling effort: (1) facilitation of human understanding and communication, (2) support for process improvement, (3) support for process management, (4) automated guidance in performing a process, and (5) automated execution support. Given our stated scope in this research study, the first three objectives are addressed.

This research paper also attempted to model a suitable technology centric approach to support decision makers in UAE and realize the vision of e-government transformation. The proposed framework was developed based on revisions of various international practices already carried out in the area of citizen-centric e-government initiatives. We refer to the framework here as CIVIC IDEA, an abbreviation for "***C****itizen* ***I****nclusive* ***V****ision realized through* ***ID*** ***C****ard* ***I****ntegrated* ***D****elivery of* ***E****-government* ***A****pplications."* The approach is envisaged to support the government of the UAE in achieving its vision of e-government transformation, while leveraging the strengths of the UAE national ID card initiative. Figure-6 summarizes the different components of the framework, each of which will be discussed in the following sections.

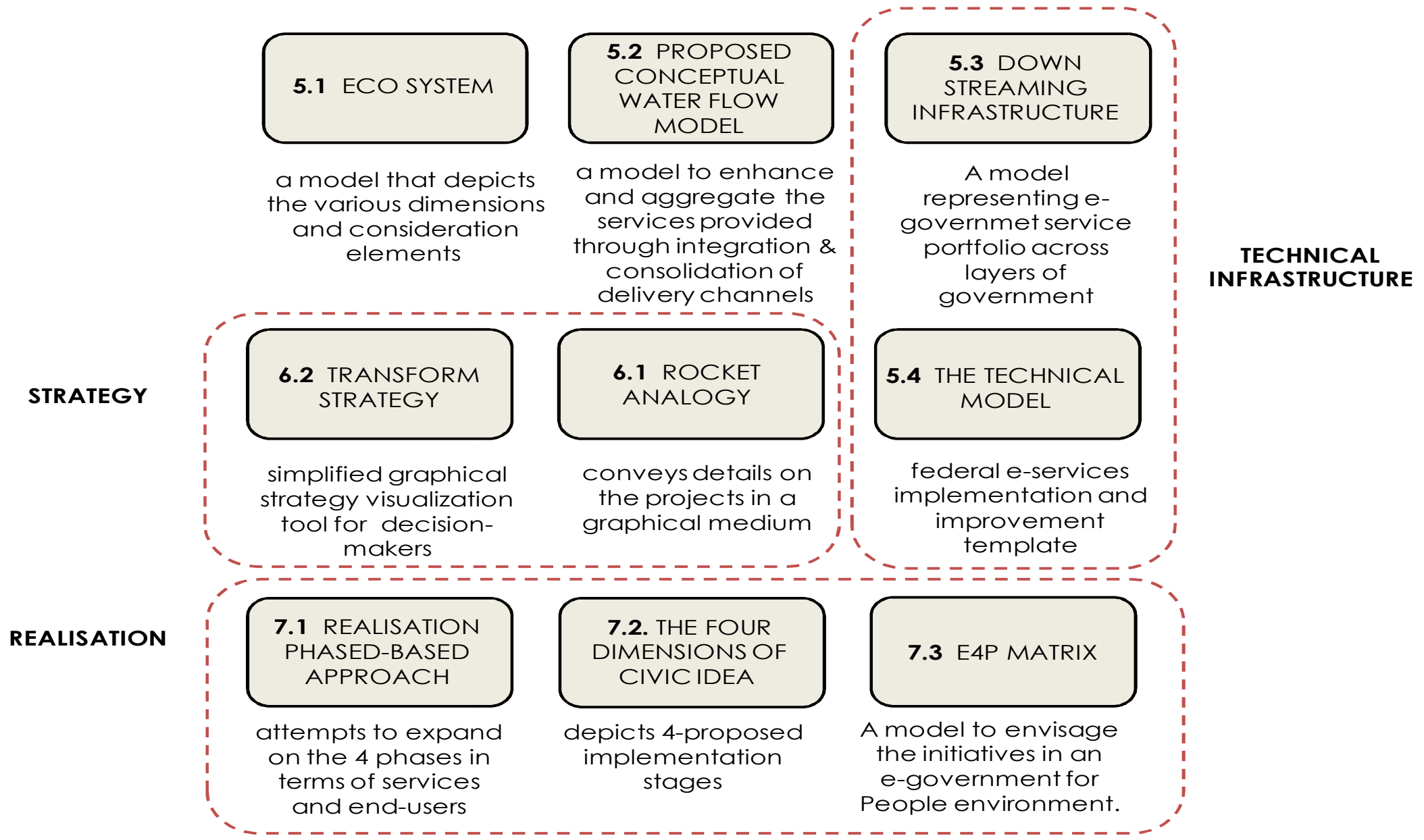

Figure-6: CIVIC IDEA Framework Components

## 5.1. CIVIC IDEA Ecosystem

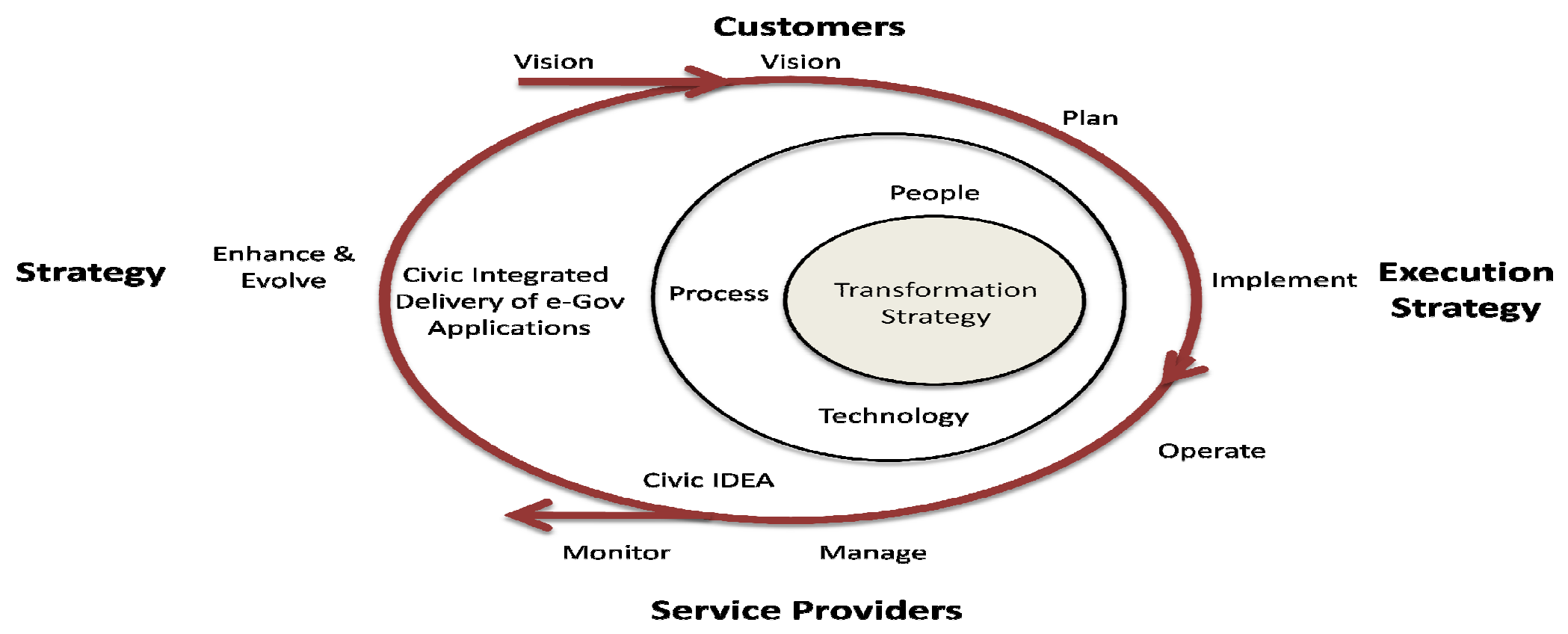

Figure-7: CIVIC IDEA Ecosystem

In an eco system of e-government, strategy and execution are equally critical. Therefore the challenge lies more in collective execution, taking into consideration the dimensions of people, processes and technology towards building an effective and integrated delivery of e-government applications. The following diagram (Figure-7) depicts the overall ecosystem for the CIVIC IDEA concept. In the diagram, we can see two sets of dimensions. The first dimension maps strategy versus execution. The key message of the above visualization is that both strategy and its execution are equally important and that neither a good strategy implemented poorly, nor a poor strategy implemented well, serves the overall objectives of e-government transformation. The other dimension in the above diagram maps service providers versus their customers. As with the former dimension, the key message of the latter dimension is also that both elements are equally important, in that innovation of new services without convenient delivery channels and tools for customer interactions, are as good as having no e-services to offer your customers. At the core of the eco system, is the transformation strategy, addressing the key elements of people, process and technology and its outer layer is comprised of citizen centric e-government applications implemented based on this new strategy. The evolution of the ecosystem consists of defining a new vision, goals, plan for the implementation of the plan, post implementation operations of the solution, monitoring of service usage and finally, the evolution of services based on the new requirements. Having explained the eco system, we will now delve into the conceptual models that form the foundation for CIVIC IDEA realization.

## 5.2. Conceptual Model

While attempting to build the solution models, it is important to have a conceptual foundation that conveys the various components of the solution. In doing so, we envisage e-government through a water flow model. In such a model, the overarching federal e-government strategy needs to be comprehended by federal and local agencies who will in turn translate these strategies into e-services for the citizens. Figure-8 illustrates the proposed conceptual model for e-government. The assumption here is that the execution based on this conceptual model will help enhance and aggregate the services offered by the service providers through integration and consolidation. Such a transformation will also require a strong focus on delivery channels to allow services to be taken to the door steps (or fingertips) of citizens.

Such a transformation would also require an increase in citizen capabilities to consume these services and enjoy their benefits. During this entire process, the government need to obtain feedback and input, in the form of Business Intelligence (BI), thereby enabling the government to fine tune its policies and strategies. Service down-streaming is one of the important foundations of CIVIC IDEA. The e-government service portfolio of the UAE consists of various layers and specializations and these services are constantly refined. However, in order to achieve uniformity across the various layers of government, it is important to have a standardized federal service template which acts as the blueprint for the implementation and improvement of e-services. See also Figure-9.

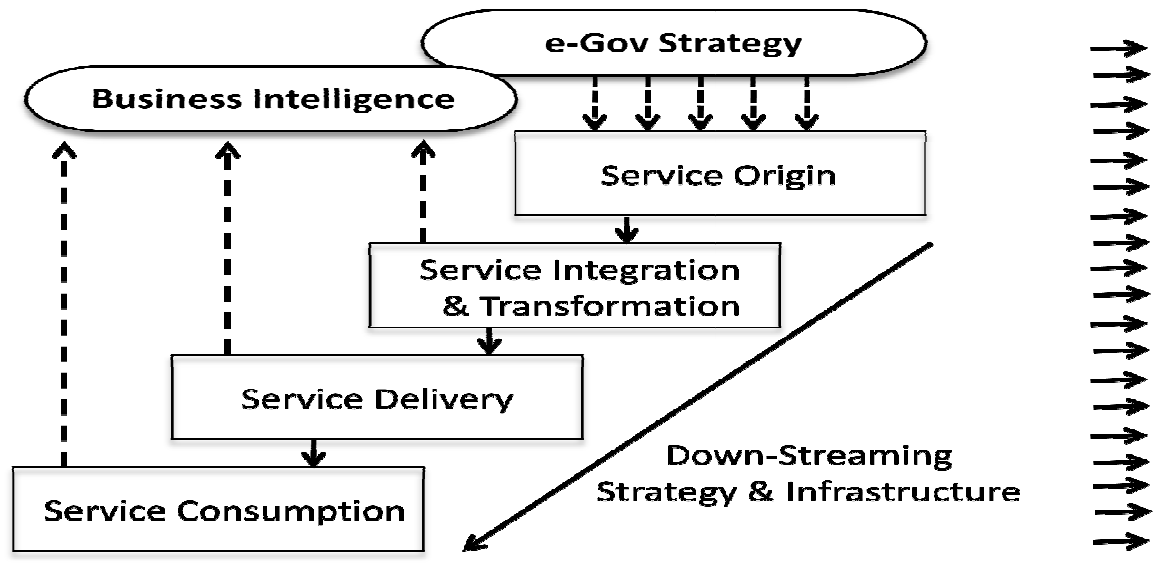

Figure-8: proposed conceptual model for e-government

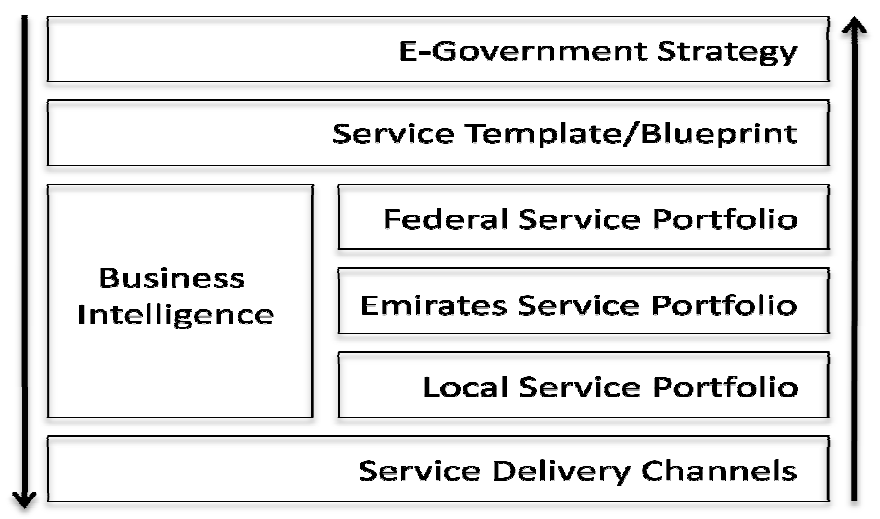

Figure-9: the “Down Streaming” infrastructure

In our opinion, successful realization of CIVIC IDEA depends on the down streaming of infrastructure and standardized service templates. These service templates would act as the blueprint for the service portfolio at federal, emirate and local levels to standardize the types of services offered. As the service template allows specialization and fine-tuning at each level, a service gets refined as it passes through many levels of specialization before it reaches the end customer via distinct delivery channels.

## 5.3. CIVIC IDEA Technical Model

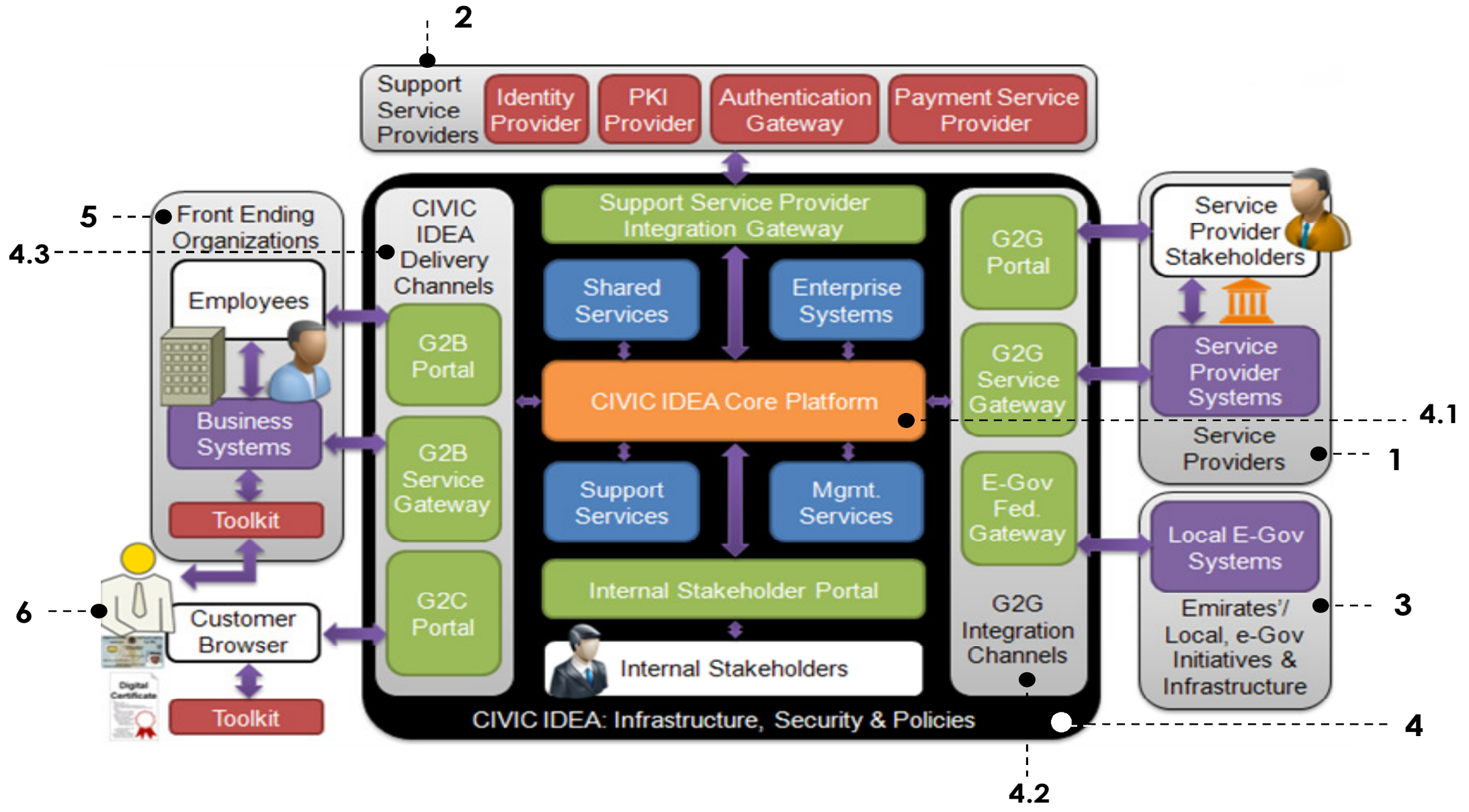

Figure-10: Enterprise level integrated view of the CIVIC IDEA

The translation of the above conceptual model into the enterprise architecture is the next step in the CIVIC IDEA realization. The development of the model took the following entities as primary design elements:

1. Service Providers
2. Support Service Providers
3. Existing E-Gov Systems
4. CIVIC IDEA Infrastructure
    - 4.1 Core Platform
    - 4.2 Integration Channels
    - 4.3 Delivery Channels
5. Front Ending Organizations
6. End Customers

Figure-10 presents the enterprise level integrated view of the CIVIC IDEA in the context of the UAE. This model envisages the technological requirements of realizing the "Down Streaming" infrastructure in the conceptual framework discussed earlier. In the following sub-sections, we will elaborate and explain the main components of the proposed CIVIC IDEA infrastructure platform.

### 5.3.1. CIVIC IDEA Core Platform

A key intention of the CIVIC IDEA platform is to enable faster and seamless integration of service providers and delivery channels. The core platform for this is envisaged to be built around a Service Oriented Architecture[4] (SOA) that leverages cloud computing[5] and virtualization[6] technologies. This would ensure the scalability and cost effectiveness of such a transformation. SOA technologies such as Enterprise Service Bus[7] (ESB) and enterprise messaging framework[8] can enable large scale integrations and communication in a standard manner. The entry points into the CIVIC IDEA platform are through standardized integration gateways and portals (highlighted in **green** in Figure-10). Such architecture would leverage reliable asynchronous messaging between service providers and the platform, whereby the synchronous real-time communication is used to ensure responsiveness. There are four categories of services (highlighted in **blue** in Figure-10) built on top of the CIVIC IDEA platform to facilitate the service integration and management, which are discussed in the following sections.

### 5.3.2. Shared Services

Shared services are value added services that can be leveraged by the service providers to achieve inter agency collaboration. Examples of shared services are audit, alert and workflow management.

### 5.3.3. Enterprise Systems

The platform is envisaged to use Enterprise Systems such as directories, databases, email and storage servers as its back-end.

### 5.3.4. Support Services

Following are the support services (highlighted in **red** in Figure-10) that would be leveraged by the CIVIC IDEA platform:

- Identity Services from identity providers such as Emirates Identity Authority to identity attribute queries;
- Public Key Infrastructure Services from nationally recognized Certificate Authorities, for certificate based authentication and digital signatures;
- Authentication Gateway services for authenticating the users transacting through CIVIC IDEA; and
- Payment Service Providers for processing fees for receipt of services.

---

[4] Service Oriented Architecture is a flexible set of design principles used during the phases of systems development and integration to support communications between services. A system based on a SOA architecture will package functionality as a suite of interoperable services that can be used within multiple separate systems from several business domains.

[5] Cloud computing is a general term for anything that involves delivering hosted services over the Internet. These services are broadly divided into three categories: Infrastructure-as-a-Service (IaaS), Platform-as-a-Service (PaaS) and Software-as-a-Service.

[6] Virtualization is the creation of a virtual (rather than actual) version of something, such as an operating system, a server, a storage device or network resources. There are three areas of IT where virtualization is making headroads, network virtualization, storage virtualization and server virtualization. Virtualization can be viewed as part of an overall trend in enterprise IT that includes autonomic computing, a scenario in which the IT environment will be able to manage itself based on perceived activity, and utility computing, in which computer processing power is seen as a utility that clients can pay for only as needed. The usual goal of virtualization is to centralize administrative tasks while improving scalability and workloads.

[7] Enterprise Service Bus is a software architecture construct which provides fundamental services for more complex architectures.

[8] Enterprise messaging framework is a set of published Enterprise-wide standards that allows organizations to send semantically precise messages between computer systems. They promote loosely coupled architectures that allow changes in the formats of messages to have minimum impact on message subscribers. EMS systems are facilitated by the use of XML messaging, SOAP and web services.

Currently, there are many initiatives in the UAE for each of these support services and the platform that can be leveraged in the transformation process.

#### 5.3.5. Management Services

Management services help to publish services from various service providers, monitor their usage and effectiveness, define and manage business processes, in addition to gaining valuable Business Intelligence (BI).

### 5.4. Standards

CIVIC IDEA assumes that the core platform should be benchmarked against global standards to ensure high degree of interoperability with open systems and commercial off the shelf (COTS) components. In general, e-government specifications must be developed based on an open integration platform and to facilitate the unification of interrelated business systems from providing applications development, operating infrastructure middleware to the application platform. Thus, and in order to meet the evolving needs of e-government, we need to use technologies complying with international technical standards in terms of policies and frameworks that facilitates interoperability between different systems. Some of the relevant and common standards are listed in Table-3 below.

Table-3: International standards

| **Platform Component** | **Standards** |
|---|---|
| Architectural standards | • Service Oriented Architecture (SOA)<br>• Enterprise Integration Patterns like Enterprise Service Bus (ESB) |
| Technology standards | • Enterprise Messaging (JMS)<br>• Web services, Federation, SAML, XACML, UDDI, etc. |
| Process standards | • Business Process Execution Language (BPEL), bXML, OASIS DSS. |
| Communication & Protocols | • SOAP, HTTPS, IPv6, etc. |
| Security Standards | • SSL v3, PKI, etc. |

The following section introduces an interesting framework developed as a part of this study, with the aim of graphically representing an overarching e-government strategy.

## 6. CIVIC IDEA: STRATEGY

The process of adopting advanced ICT solutions for the transformation of e-government faces many challenges. Due to the complex nature of these projects and the sheer number of stakeholder's involved, effective visualization and management of such initiatives is highly critical but needs to be simple in order to accelerate understanding off and buy in into the framework. However it is important that the framework represent all important aspects of the e-government strategy. That said, despite years of governmental efforts to implement e-government initiatives, there are no commonly established methods and frameworks for the visualization of an overarching e-government strategy. A comprehensive framework needs to account for how the different supporting and impeding forces impacting projects being implemented as a part of such a strategy. Given that such a framework will also have a long lifecycle and encompass a broad scope, the framework also needs to remain applicable regardless of changes in the environment. As most projects go through many iterations of technical and process changes, any changes within the ecosystem should not risk the validity of the strategy framework. Hence the framework needs to be adaptable to changing environments and should be defined in a technology neutral manner. Such an approach will also allow the framework to act as the bridge between decision makers and implementers, thus reducing the mismatch between the expected versus realized outcomes.

## 6.1. Rocket Analogy

A good analogy which is well understood by key decision makers can convey more information than lengthy text based description (see also Figure-11). However it is important to note that an analogy cannot replace the formal definition of a strategy, but can only be used as a means of easily conveying the key messages. Primarily the analogy selected should be able to convey maximum details about the project being considered through a graphical medium so as to save time as well as enable better coordination and reduce ambiguity. Looking at the below diagram we can identify the forces that act on the rocket. Inertia is an opposing force that is commonly encountered in a project requiring change. In order to overcome inertia, one needs to apply heavy thrust till the rocket (project) gains significant momentum. Once in motion (execution) the rocket faces continuous opposing forces which though not as strong as the inertia, can still slow down the projects or take it off course. These resistances can be in the form of coordination issues, technical issues, lack of standards, etc. For a rocket to overcome these opposing forces, thrust must be applied. The thrust can come as a push from the management or pull from the customer side. During the course of flight there is a need to continuously monitor the flight path to detect any deviations. These deviations once identified need to be communicated to the rocket navigation system to take controlling actions.

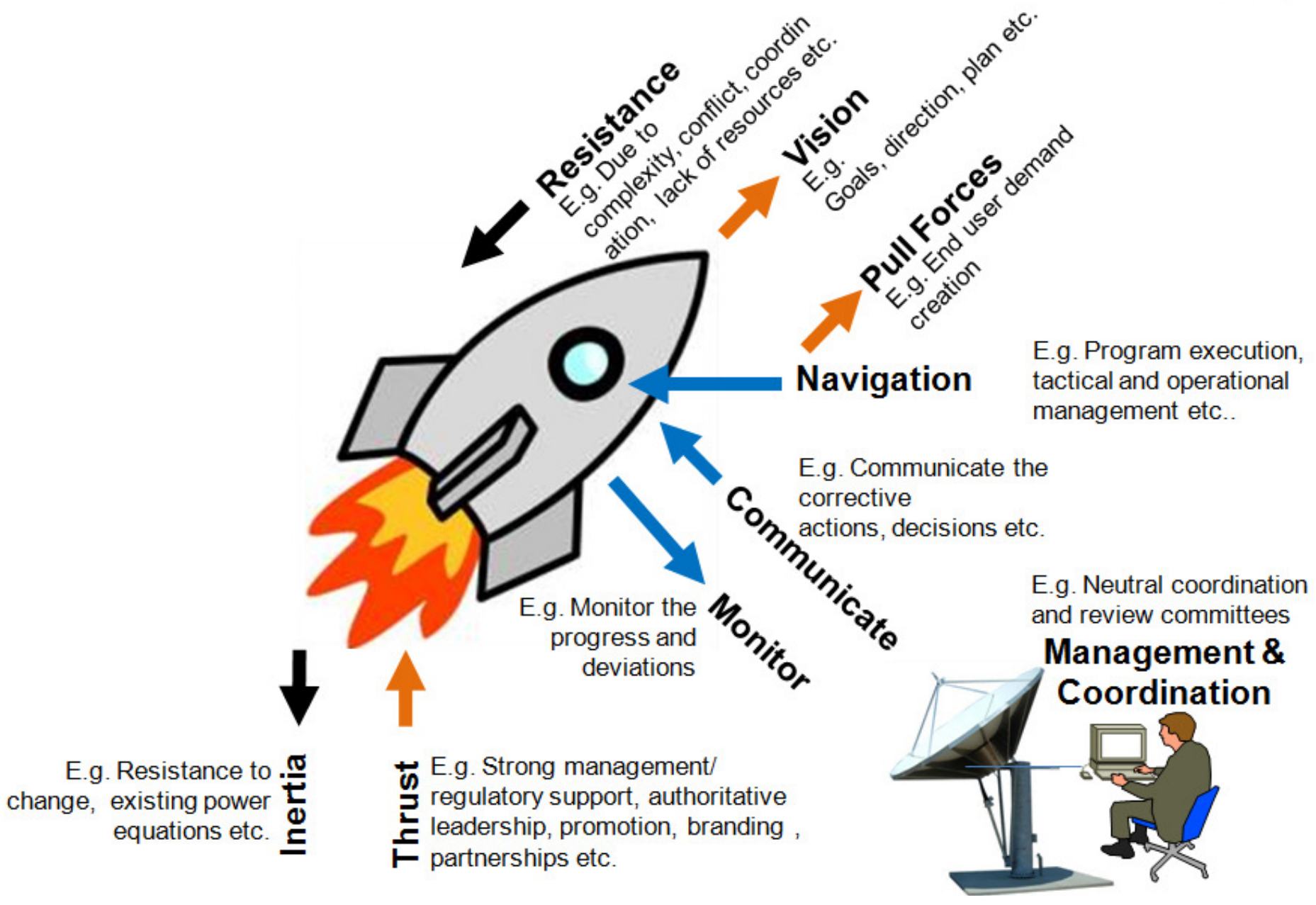

Figures-11: Rocket Analogy

## 6.2. TRANSFORM Strategy

From the analogy described above, we derived a model that maps to the e-government domain artifacts and problem statements. This model named as **T**hrust, **R**esistance **A**nd **N**avigation **S**trategy **Form** or in short as **TRANSFORM** is illustrated in Figure-12. Such graphical visualization of the strategy is likely to be beneficial to decision makers, as it provides a non-technical visualization. It provides a simplified yet a comprehensive conceptualization of what e-government strategy is all about. Table-4 provides a short highlight on each of the key focus areas within the TRANSFORM strategy.

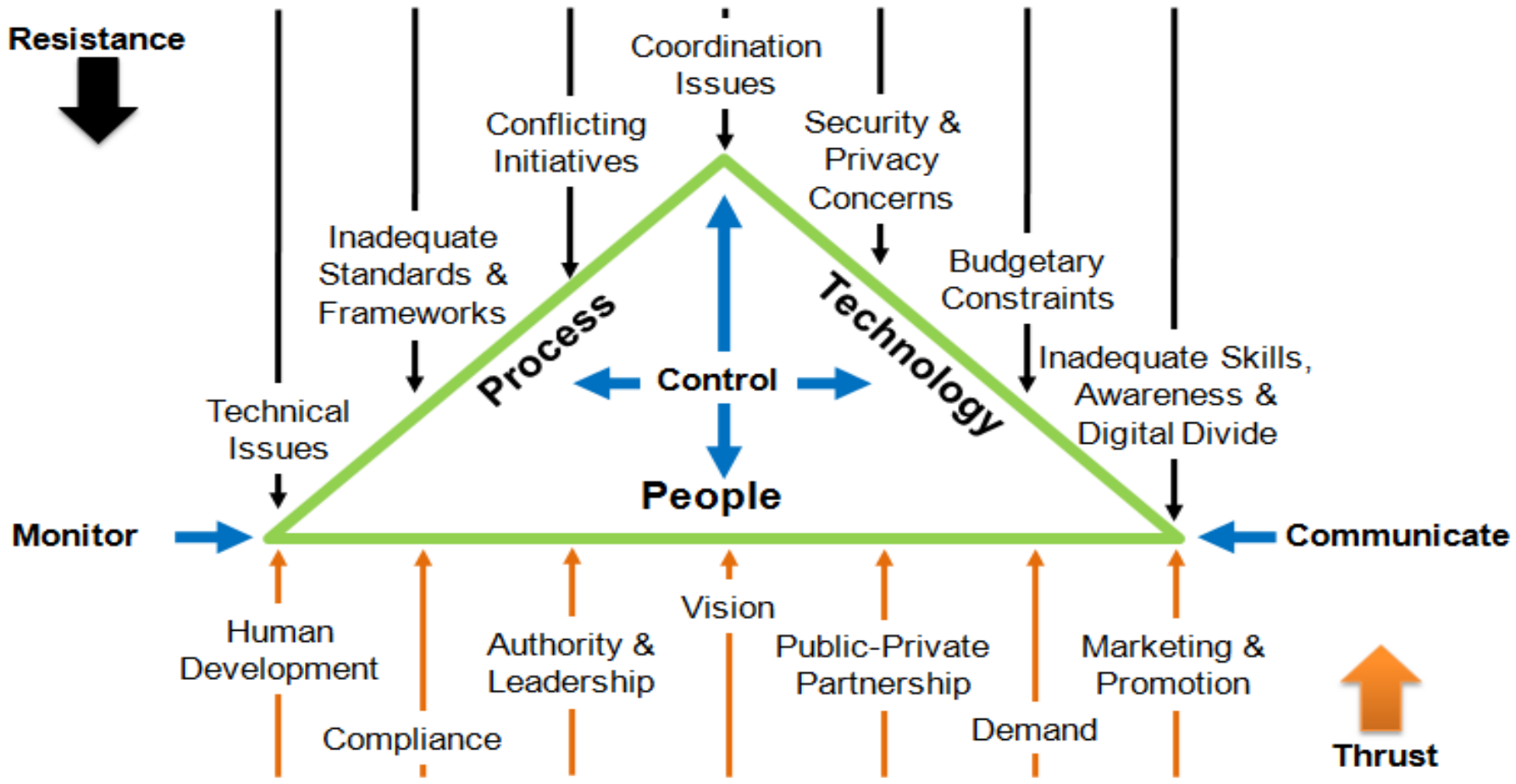

Figure-12: TRSNAFORM Strategy

Table 4: Major Resistances and Thrust areas in e-Government

| **Resistance** | **Thrust:** |
|---|---|
| *Coordination Issues:* Variations in legal, regulatory and administrative regimes on different sides of these boundaries can inhibit and block the flow of information and services. | *Vision:* quantifiable strategic outcomes, periodic reviews and progress assessment. |
| *Budgetary Constraints:* Difficultly in quantifying and measuring the cost/benefits of e-government initiatives**.** | *Authority and Leadership:* lack of authority and leadership |
| *Digital Divide:* Social and economic divides – demarcated by wealth, age, gender, disability, language, culture, geographical location, size of business and other factors**.** | *Demand:* lack of perceived benefits resulting in inadequate motivation for citizens to avail e-government services. |
| *Security and Privacy Concerns:*security and privacy individual's data and the risk of information and identity theft | *Public Private Partnership:* poor definition of policies enabling such partnerships |
| *Technical Issues & Inadequate Standards:* inappropriate user interfaces to e-government systems and interoperability issues. | *Human Development:* Inadequate skilled resources |
| *Resistance to Change:* inadequate staff skills, lack of training and investment in enhancement of ICT knowledge and the fear of change. | *Marketing and Promotion:* Lack of marketing and branding strategies to gain wide visibility, recognition and demand. |
| *Conflicting Initiatives:* competition between initiatives to achieve similar outcomes. | *Compliance:* lack of common standards, agreed procedures and methodologies, e.g., legal and regulatory policies and guidelines as well as technical and operational standards |

To understand how we propose to define the navigation strategy for our framework, let us first revisit Figure-7, which illustrates an e-government project as an ecosystem of people, processes and technologies. These three components work in close coordination in the implementation of any project. To ensure their alignment with common goals and meeting defined performance criteria, it is a common practice to have an independent *review committee* which periodically monitors and reviews the progress of each project. Observations from such reviews can then be communicated to the project leadership to enact specific controls to implement any necessary corrections and re-alignment. The proposed TRANSFORM strategy framework presented here is a visual tool that represents the e-government projects in a technology neutral and abstract manner, using an analogy that is widely familiar and simple. This should enable strategic decision makers in seeing through the challenges faced by the initiatives and provide them with the necessary thrust needed to overcome their challenges. This can also greatly bridge the gap between policy makers and implementers, as a common representation of the projects resulting in higher clarity and reduced misalignment.

## 7. CIVIC IDEA: REALIZATION APPROACH

The earlier sections in this study discussed the models and strategy for the realization of the proposed framework of CIVIC IDEA. However such large scale transformation cannot be achieved in one go and needs to be deployed in phases, with each phase initiated based on the successful achievement of outcomes in earlier phases.

### 7.1. Stages of CIVIC IDEA Realization

There are four key stages to realizing CIVIC IDEA. The following diagram (Figure-13) depicts four key focus areas to drive through the four phases, with each phase trying to expand the coverage in terms of services and end users. In summary, the **enable** phase is more of a preparatory phase where the foundation for transformation is laid. The **enhance** phase is used to develop blue prints, standards and basic infrastructure. Having created the basic infrastructure, smart projects are initiated in the **establish** phase with the aim of gaining wider support, increasing visibility and creating demand. All through the first three phases, the overall strategy gets refined and is now ready for **expansion** to reach maximum coverage.

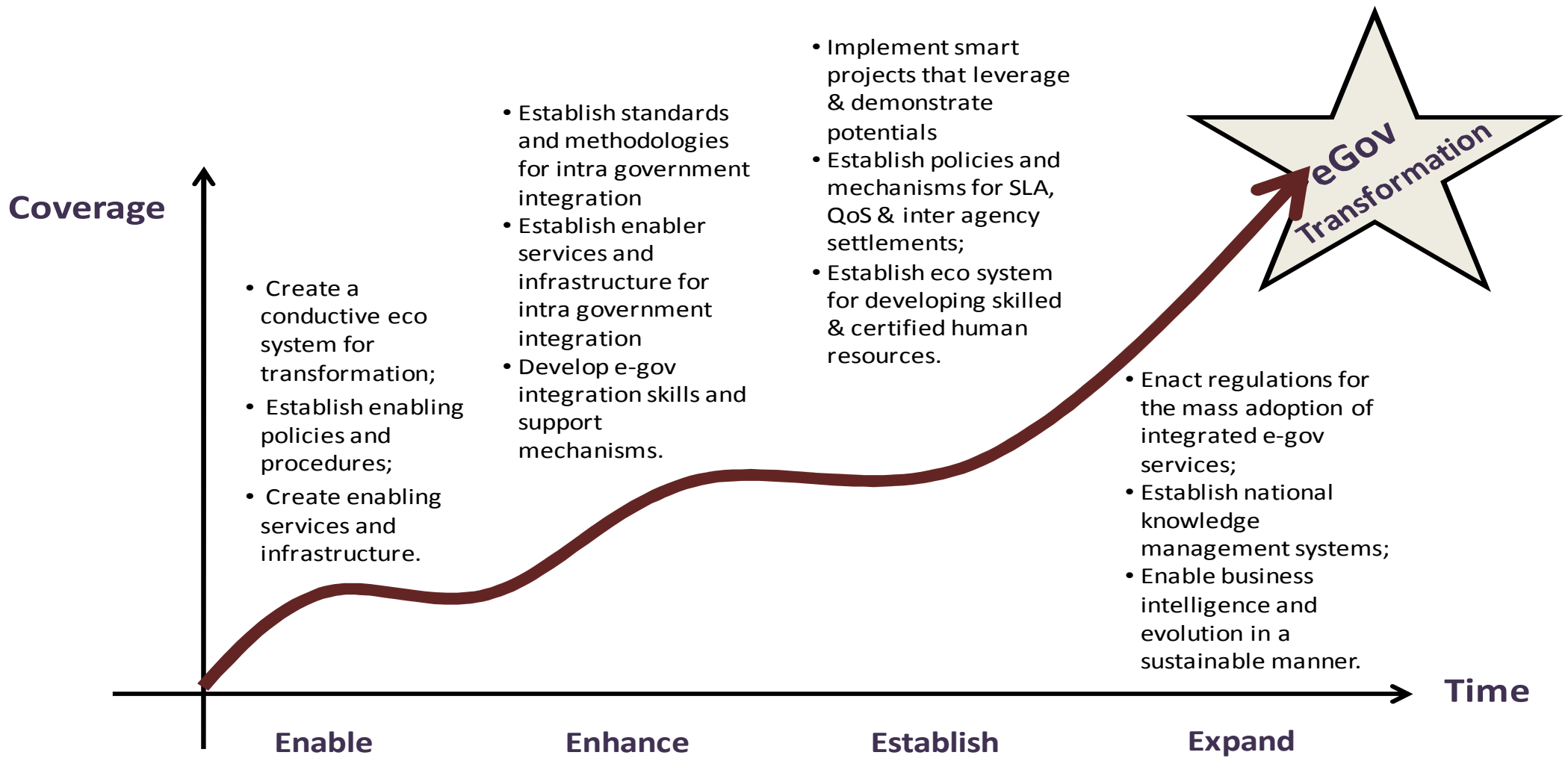

Figure -13: Driving through CIVIC IDEA

## 7.2. Dimensions of CIVIC IDEA

The stages in the CIVIC IDEA realization define the stages in the timeline. The following diagram (Table-5) lists the four dimensions of CIVIC IDEA with core focuses. At each of the stages, we need to achieve higher maturity in each dimension of Policies, Processes, Projects and People.

Table 5: The four dimensions of CIVIC IDEA

| **Policies** | **Processes** |
|---|---|
| • e-Gov authority and leadership;<br>• regulatory acts and laws towards enforcement;<br>• public private partnership. | • Standards & guidelines;<br>• Reviews & coordination mechanisms;<br>• Project management & PKI monitoring. |
| **Projects** | **People** |
| • Specific project covering the areas of infrastructure building, solutions, communications, delivery channels, service access points, supporting mechanisms, etc. | • Human resource development towards e-Gov resources;<br>• Awareness, promotion, marketing, branding. |

## 7.3. Profile of Initiatives

Based on the realization stages and dimensions we then arrive at a model to envisage the initiatives in an **E4P** (this also represents **E-government** for **P**eople) matrix which is shown in Figure-14 below.

| | Enable | Enhance | Establish | Expand |
|---|---|---|---|---|
| **Policy** | ▪ Cyber laws<br>▪ Digital signature act<br>▪ Security & privacy act<br>▪ Vision<br>▪ Goals | ▪ Performance Management Policy<br>▪ Change management & Prioritization<br>▪ National ID Act | ▪ Public Private Partnership Policy | ▪ BI driven continuous policy improvements |
| **Process** | ▪ E-Gov Authority<br>▪ Expert committees<br>▪ Coordination & collaboration processes<br>▪ Federal Enterprise Architecture (FEA) | ▪ Standards & Business Processes<br>▪ E-Gov Interoperability framework<br>▪ KPI Management<br>▪ Review & monitoring<br>▪ Project coordination & management | ▪ Services directory<br>▪ Revenue settlement<br>▪ SLA management<br>▪ Customer/ Citizen Surveys & Feedbacks<br>▪ Business Intelligence (BI) | ▪ BI driven process improvements |
| **Projects** | ▪ PKI & FIM<br>▪ ID Card Toolkit<br>▪ Toolkit distribution & integration support<br>▪ FIM integration support | ▪ CIVIC IDEA platform<br>▪ CRM & support infrastructure<br>▪ Develop integrated delivery channels | ▪ Electronic Benefit Transfer (EBT)<br>▪ Online Delegation<br>▪ Enhance access points<br>▪ Integrate/Migrate existing e-Services to CIVIC IDEA | ▪ Coverage in terms of reach and services<br>▪ Exporting solutions |
| **People** | ▪ Promotion<br>▪ Awareness<br>▪ Branding | ▪ Integration skills development<br>▪ E-Gov support skills<br>▪ ICT literacy<br>▪ CIVIC IDEA Conference | ▪ E-Gov development skills<br>▪ Certification programs | ▪ Knowledge management<br>▪ Exporting consultancy services |

* The initiatives highlighted in red are already undertaken by UAE.

Figure-14: The e-government for people matrix

## 8. KEY THOUGHTS AND CONSIDERATIONS

Having presented the proposed framework, this section outlines some key thoughts, considerations and recommendations for practitioners in the field of e-government. Together they are meant to raise awareness, aid the building of resilient plans, and enable the "endogenization" of institutions and the creation of a favorable implementation environment.

### 8.1. Key Success Factors of Citizen Centric E-government

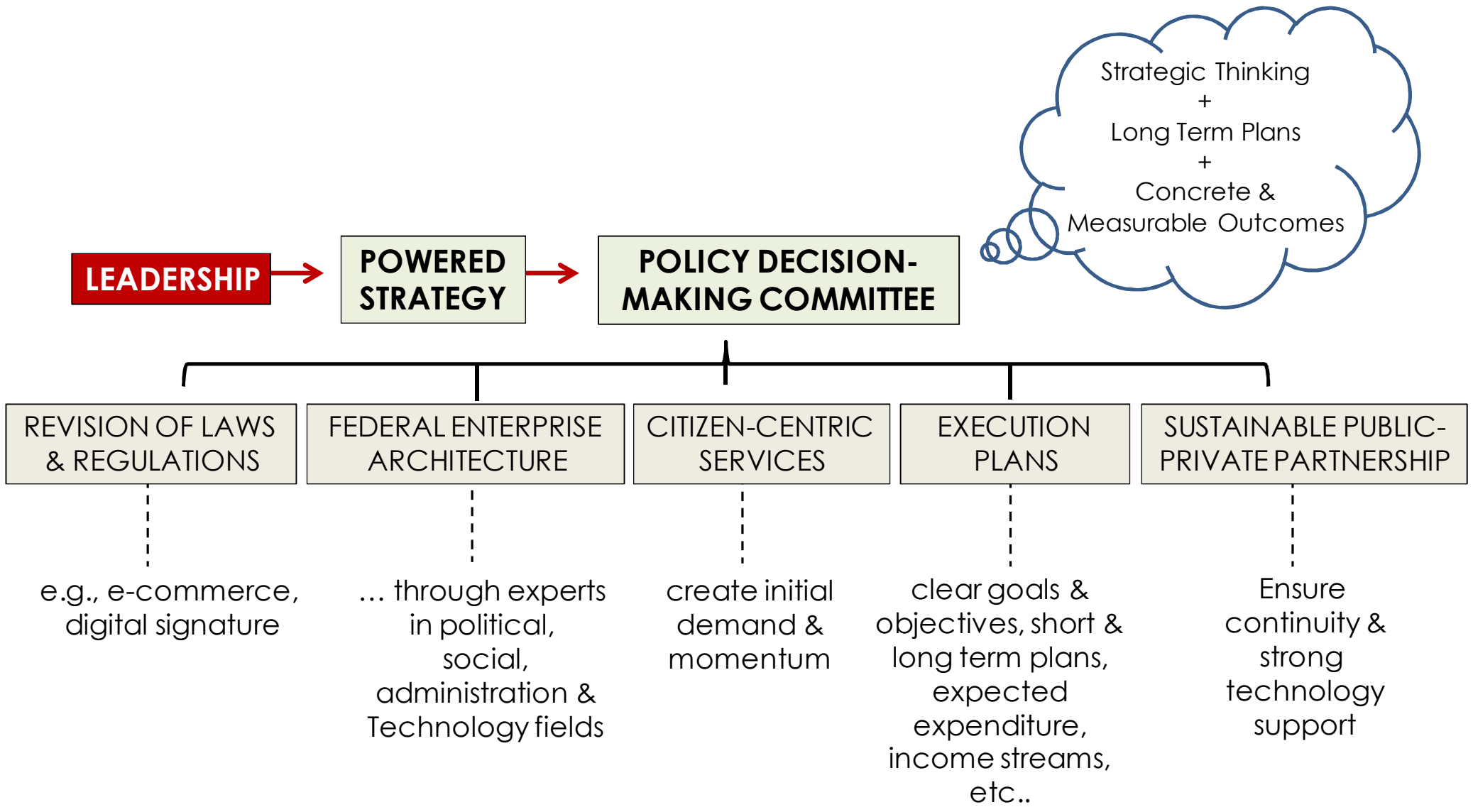

Figure-15: Key success factors for citizen centric e-government

Success of citizen centric government depends on many factors as depicted in Figure 15. Primarily, strong e-government leadership is essential for uniform and centralized decision making on e-government initiatives [33]. Such an approach should consist of a high powered strategy and policy decision making committee with strategic thinking and sustainable long term plans. These plans should encompass aligned e-government projects, each having concrete and *measurable outcomes*. E-government leadership should also be enabled by revision of laws and regulations e.g. e-commerce and digital signature acts, in line with government process reforms [34-36]. Adoption of federal enterprise architecture arrived at with the help of experts in political, social, administration and technology fields, is also vital to the success of e-government initiatives [35,37-38]. Engaging in these activities would provide the necessary direction and structured thinking necessary to launch programs that complement one another in achieving national goals. On the flip side, citizen demand for e-government services is essential for the success of e-government initiatives. Hence e-government initiatives should ensure citizen centric e-services which can be accessed through convenient channels of delivery. Governments need to look at ways to create the initial demand and momentum, through services that delivers direct benefit to citizens. Such services can be in the field of health, education, social affairs (e.g. subsidies and pension) etc. where the citizen can see a tangible benefit which in turn spurs demand for e-services.

E-government projects and programs should be performance oriented with measurable outcomes [39-41]. Clear goals, objectives, short and long-term plans, with expected expenditure, income streams and deadlines are some of the attributes that should be defined for such projects [42] and performance criteria should include both qualitative and quantitative measures. Sustenance is the last word in e-government. Hence e-government leadership should provide an ecosystem for sustainable public private partnership, which is revenue generating and evolving to ensure continuity and strong technology support for the envisaged initiatives. That all said, the following sub-section presents some key thoughts around improvement opportunities in the field of e-government in the United Arab Emirates.

### 8.2. Key Improvement Areas for UAE

**Branding** is an important aspect of the UAE's e-government strategy that needs further attention. Though the UAE has undertaken numerous steps in the field of e-government, its efforts needs to be documented in a consolidated knowledge base and disseminated via for example presentations at international conferences and publications in leading industry journals. Such steps would allow the UAE to gain broader visibility for its work, increased opportunities for peer reviews and in turn feedback and input from experts in this field. Another area for improvement is **Service Coverage**. There is an imminent need to create integrated and shared services that would demonstrate the potential of transformed government and thus leads to demand creation. Coverage should ensure that all services within each line of businesses are covered and that there is also coverage across lines of business. Another dimension that should be addressed here is the inclusiveness of access to ensure that all the stakeholders are uniformly covered.

There is also a need to consolidate e-government initiatives and increase focus on **enhancing quality of life**. This could be achieved through provision of convenient access channels that are accessible around the clock or via personalized e-government portals. Such changes will help accelerate citizen usage of government services as citizens will now be able to access services via fast and convenient methods and via reduced effort and time investment. A prerequisite for such uniform access to e-services would be unified identification and authentication. Having implemented a smartcard based national identity scheme, UAE should leverage it as a medium for citizen identification and for citizens to access all the aforementioned e-services. However, successful realization of unified e-government initiatives needs **resources and capacity building**. The UN e-government index highlights the lack of skilled manpower to implement and operate technology intensive e-government initiatives. A prerequisite is the requirement for technology literacy among citizens, so as to maximize the benefits off and fully leverage e-government services. E-government efforts result in tremendous knowledge creation and consolidation and require centralized coordination and management so that all stakeholders can access and leverage the insights and experiences of one another. Having said this, the next section concludes this research report.

## 9. CONCLUSIONS

Governments around the world have pursed e-government programs seeking to electronically govern internal and external operations and to provide coherence between the various administrative government units so that they work to complement and complete each other. However, and despite the fact that many governments have injected substantial investments, most e-government initiatives in our view have not delivered the transformation environment sought from their implementation. This research study was developed to support the United Arab Emirates in pursuing its objective towards e-government transformation. It presented an innovative framework developed from a government practitioner's viewpoint and in light of the existing literature in the field. The recommended approach is an amalgamation of learnings from various e-governments initiatives across the globe. The presented framework in this research study was particularly designed to support decision makers and present them with key information and focus areas in e-government initiatives. The framework proposed incorporates some significant conceptual models to enhance leadership understanding and their ability to respond to challenges. It defines a comprehensive approach addressing technology, strategy and the broader approach to realizing e-government transformation. It proposes many innovative models to support the visualization of numerous dimensions of transformed e-government. In the hands of strong e-government leadership, this is envisaged to act as a powerful tool to communicate and coordinate initiatives. Assessment of the success of this proposed framework was beyond the scope of this research study. Certainly, further research and application is needed to examine the practicality of the proposed framework and its components and the mechanics by which it may be practiced. Last but not least, the maturity of e-government requires significant efforts by both practitioners and researchers to support the development of horizontal and vertical e-government integration. From this standpoint, this research study attempted to make a contribution in this critical and imperative area of knowledge and practice.

## ACKNOWLEDGEMENTS

The work presented in this study is an extension to a previous research study published in the Journal of Computer Science and Network Security, under the title: "A strategy framework for the risk assessment and mitigation for large e-Government project." Vol. 10 No. 10, November 2010. The author would like to thank Mr. Dennis DeWilde and Mr. Adeel Kheiri from Oliver Wyman for their feedback on this study, and their assistance in improving the overall structure and quality of the article.

**Dr. Ali M. Al-Khouri**

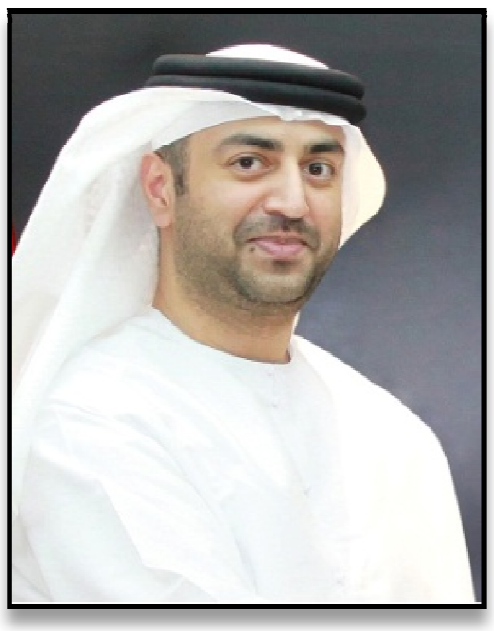

Dr. Al-Khouri holds an Engineering Doctorate (EngD) in the field of large scale and strategic government programs management from Warwick University. He is currently working with Emirates Identity Authority as the Director General. During the past 20 years, he has been involved in many strategic and large scale government programs. He is been an active researcher in the field of revolutionary developments in government context and has published more than 30 articles in the last 4 years. His recent research areas focus on developing best practices in public sector management and the development of information societies with particular attention to e-government applications.